# Oxide perovskite BaSnO$_3$: A promising high-temperature thermoelectric material for transparent conducting oxides

Xiefei Song [*,a], Guangzhao Wang [b], Li Zhou [a], Haiyan Yang [a], Xiaopan Li [a], Haitao Yang [a], Yuncheng Shen [a], Guangyang Xu [a], Yuhui Luo [*,a] and Ning Wang [*,c]

[a] College of Physics and Information Engineering, Zhaotong University, Zhaotong 657000, Yunnan, China

[b] Key Laboratory of Extraordinary Bond Engineering and Advanced Materials Technology of Chongqing, School of Electronic Information Engineering, Yangtze Normal University, Chongqing 408100, China

[c] School of Science, Key Laboratory of High Performance Scientific Computation, Xihua University, Chengdu 610039, Sichuan, China

[*] Corresponding author.
E-mail: xiefsong2023@163.com (Xiefei Song), lyh277@163.com (Yuhui Luo), ningwang0213@163.com (Ning Wang)

ABSTRACT: The new technology of energy conversion must be developed to ensure energy sustainability. Thermoelectric (TE) materials provide an effective means to solve the energy crisis. As a potential TE candidate, the TE properties of perovskite have received extensively attention. We here investigate the TE transport properties of the transparent conducting oxide (TCO) $BaSnO_3$ by first-principles calculations. We find that the $BaSnO_3$ perovskite exhibits outstanding dynamic and thermal stabilities, which provide excellent electronic and thermal transport properties simultaneously. These properties contribute to the remarkable Seebeck coefficient and power factor, which gives rise to the *ZT* of *n*-0.37 and *p*-1.52 at 900 K. Additionally, doping and nanostructure open prospects for effectively improving the TE properties of $BaSnO_3$. Our work provides a basis for further optimizing the TE transport properties of cubic $BaSnO_3$ and may have worthwhile practical significance for applying cubic perovskite to the high-temperature thermoelectric field.

# 1. Introduction

It is an outstanding challenge for solving the energy crisis to find materials that achieve energy conversion and sustainable development. Recently, thermoelectric (TE) materials are of interest in directly converting energy between waste heat and electricity without any pollutants, shedding light on an effective strategy for energy dilemma. For a TE material, its TE transport properties are determined by the Seebeck coefficient ($S$), electrical conductivity ($\sigma$), thermal conductivity ($\kappa$), and temperature ($T$), due to the conversion efficiency is characterized by the figure of merit: $ZT=S^2\sigma T/\kappa$. An ideal TE material has the characteristic of $ZT \geqslant 1$ [1]. However, as is well known, the TE properties of numerous materials do not meet this condition. Specifically, as carrier concentration increases, electrical conductivity and thermal conductivity increase whereas the Seebeck coefficient decreases. Therefore, it is a great challenge to find an ideal TE material. In a word, the strong coupling of these coefficients is a challenging bottleneck for further advances.

To achieve high $ZT$ by balancing power factor ($PF=S^2\sigma$) and thermal conductivity, theoretical and experimental works have been powerful driving in the TE field. Recently, several works demonstrated that heavy doping, nanostructures, and band structure engineering nontrivially improve the TE performance [2-6]. For example, the decrease of the dimensionality for materials effectively reduces the lattice thermal conductivity (LTC) $\kappa_l$ [7], whereas the adjustment of the carrier concentration achieves a high $PF$ in heavily doped semiconductors. In addition to optimizing existing materials, the search for new TE materials is still in progress which have outstanding TE performance. Example includes those lead-free perovskites of the general formula ABX$_3$ (A=Ca, Ba, Cs; B=Zr, Sn; X=I, Br, S, Se) [1,8-12], as promising thermoelectrics, have attracted significant interest because of their intrinsically high Seebeck coefficient and low thermal conductivity. However, these materials typically exhibit poor stability despite their strong environmental friendliness.

Lead-free oxide perovskite materials have been extensively studied in the application of photocatalysis, electrocatalysis, and photovoltaics [13-16] owing to their extraordinary optoelectronic properties. Being an *n*-type semiconductor [17-19], BaSnO$_3$ has been widely applied in the transparent conducting oxide (TCO) layer for solar cells due to their inherent properties. These properties include high electronic mobility, outstanding electrical conductivity, and oxygen stability [20-24]. Additionally, being an alternative promising TE material for high-temperature utilization [25], BaSnO$_3$ has attracted considerable attention due to its benefits, including nontoxicity, affordability, and good thermal stability. In 2017, a work demonstrated that the *n*-type doping improves the electrical conductivity of BaSnO$_3$ [26], indicating that the adjustment of the carrier concentration is an effective way to enhance the TE properties of BaSnO$_3$. Recently, theoretical studies suggested that compared to other TCO materials (such as CdO, SnO$_2$, and ZnO), BaSnO$_3$ is an ideal TE material due to its low thermal conductivity [27]. In addition, the experimental investigations [19] reported that BaSn$_{1-x}$Sb$_x$O$_3$ (x=0, 1, 5, and 10 mM) has a high power factor of 2.1 µW/mK$^2$ and a remarkable Seebeck coefficient of 58.9 µV/K at 820 K. A work further suggested that co-doping BaSnO$_3$ perovskite with Sr and Sb has led to a high *PF* of 25.3 µW/mK$^2$ at 578 K [28]. Furthermore, theoretical works demonstrated that chalcogenide perovskite (*i.e.*, $CaZrSe_3$, $CaZrS_3$, and $BaZrS_3$) exhibit excellent thermoelectric properties [1,8-11]. Because O and S, Ba and Ca are of the same main group, one expects the outstanding TE properties of $BaSnO_3$ compared with those of $CaZrSe_3$ and $BaZrS_3$. These works mentioned above raise the natural issue: can $BaSnO_3$ be an ideal thermoelectric candidate with promising performance?

Notice that oxide perovskite $BaSnO_3$ with cubic structure is synthesized via a low-cost preparation methods of solid-state reaction [29-32], which promotes the exploration of TE properties. It is noteworthy that typical TE materials are easily oxidized in air, making them unsuitable for high-temperature applications, *i.e.*, PbTe, $Bi_2Te_3$, and $Sb_2Te_3$ [19,28]. We now pose a challenging question: can $BaSnO_3$ be applied in high temperatures? Hence, for $BaSnO_3$ to function at high temperatures, it is required to systematically explore its thermoelectric transport properties. In this work, we explored the TE properties of $BaSnO_3$ by employing first-principles calculations, which has important practical significance for related investigations in high-temperature thermoelectric fields.

## 2. Computational details

All first-principles calculations were implemented using VASP code [33] with Perdew-Burke-Ernzerhof (PBE) functional [34]. The PBEsol functional [35] and HSE06 hybrid functional [36] were applied for the calculations of structural optimization and the electronic structures, respectively. An energy cut-off of 450 eV and a *k*-point grid of 5 × 5 × 5 was utilized to ensure accuracy, and a high *k*-point grid of 13 × 13 × 13 was employed to obtain accurate TE transport properties. A convergence criterion of $10^{-3}$ eV/Å and $10^{-5}$ eV were used for the force and energy in calculations, respectively. Based on the HSE results obtained from VASP, the TE coefficients were calculated using the BoltzTraP2 code [37].

AIMD [39] simulations were performed to identify the thermal stability under different temperatures for $BaSnO_3$, as implemented in VASP. The temperature conditions were maintained via a Nosé–Hoover thermostat by using a canonical ensemble (NVT), and all simulation runs were 5 ps in duration. Phonon transport properties and LTC were performed with Phonopy code [40] and Phono3py code [41], respectively. The second- and third-order force constants were calculated by employing a 3 × 3 × 3 supercell (135 atoms) of the cubic unit cell, in which an atomic displacement of 0.0075 Å was used for the calculation of second-order force constants [42], while all three-atom interactions are considered in third-order force constants. A convergence criterion for $q$-point sampling mesh of 30 × 30 × 30 was used in the LTC calculations.

## 3. Results and discussion

### 3.1 Geometrical structures and electronic structures

The oxide perovskite $BaSnO_3$ shows a cubic structure in the space group $Pm\bar{3}m$ (No.221) with five atoms per primitive cell (Fig. 1(a) and 1(b)). From another viewpoint, $BaSnO_3$ has a isotropic three-dimensional structure. It consists of the corner-sharing octahedrons $[SnO_6]^{8-}$ connected with each other in $a$, $b$, and $c$ directions, and $Ba^{2+}$ ions are filled in the gaps. This isotropic lattice structure determines the isotropic transport properties of the BaSnO$_3$. The optimized lattice constant of 4.12 Å ($a = b = c$) agrees with the experimental data (4.12 Å) [31,32] and other calculation results (4.13 Å) [42,43].

Significantly, the precision of the band gap determines the TE performance. Fig. 1(c) and 1(d) display the band structure and the density of states (DOS) of BaSnO$_3$. The band structure shows that the valence band maximum (VBM) locates at $R$ point and the conduction band minimum (CBM) locates at $\Gamma$ point, indicating that it is of the band gap of the indirect character. The band gap of the BaSnO$_3$ we obtained is 2.46 eV, which is consistent with previous reports (2.48 and 2.40 eV) [42,43]. Additionally, the contributions from different atoms to DOS are well determined in Fig. 1(d). It has been revealed that O 2p states largely contribute to the total DOS in the VBM, and the CBM is predominately contributed by Sn 5s orbitals. Specifically, from Fig. 1(d), the total density of states can be divided into four different regions. The first region can be indexed to the electronic states located below −3 eV in the lower part of the valence band. These states mainly consist of O 2p and Sn 5p states. The second region can be indexed to the electronic states located between −3 eV and 0 eV, which are derived from the O 2p states. The third region can be indexed to the electronic states located between 2.46 eV and 4 eV of the conduction band. The electronic states in this region could be mainly derived from Sn 5s states, while O 2s and O 2p orbitals have minor contributions to the electronic states. The last region is located above 4 eV, which is predominantly composed by Sn 5s states along with a minor contribution from O 2p state. These results are consistent with the previous report [27]. It is worth noting that, from band structure, there are two valence band valleys (*i.e.*, VBM and VB1) around the Fermi level with an energy difference of 0.44 eV (as shown in Fig. 1(c)). Such an energy difference leads to a large slope of DOS distribution near the valence band maximum (as shown in Fig. 1(d)). Moreover, two conduction band valleys are also observed, *i.e.*, CBM and CB1. The energy difference of 4.36 eV between CBM and CB1 results in a small slope of DOS distribution around the conduction band minimum. Based on the band structure, we extracted effective masses ($m^*$) of 1.34 for holes and 0.14 for electrons. The differences in the band curvatures between the top of valence bands and the bottom of conduction lead to the different $m^*$ between holes and electrons. Specifically, compared with the top of valence bands, the bottom of conduction bands possesses larger band curvatures, indicating a lower $m^*$ for the electron. Furthermore, as $BaSnO_3$ has a cubic structure, we find that all the $m^*$ are nearly isotropic along all three crystal directions, which is consistent with a previous report [43] and indicates that the thermoelectric coefficients of $BaSnO_3$ will also exhibit the trends of isotropic.

**3.2 Stability and phonon structure**

Oxide TE materials are known for their excellent thermal stability that contributes to the thermoelectric performance of materials and devices. Here, to explore the thermal stability of $BaSnO_3$ under high temperatures, AIMD simulations are implemented. It is clear that the free energy remains nearly constant at high temperatures as the simulation time increases (Fig. 2a). Additionally, the cubic $BaSnO_3$ maintains structural stability during the simulation with the temperature from 300 to 900 K (Fig. 2(b)-(d)), suggesting the excellent thermal stability of this material at high temperatures.

To provide insights into the dynamic stability and phonon transport mechanisms of BaSnO$_3$, we perform the calculations of the phonon band and DOS. Fig. 2(e) shows that the phonon spectrum has 15 phonon modes with three acoustic modes. Namely, transverse acoustic (TA), longitudinal acoustic (LA), and flexural (ZA) modes are highlighted by black, red, and green lines, respectively. It should be noted that there are no imaginary frequencies in the phonon spectrum, indicating that the cubic BaSnO$_3$ is dynamically stable. The phononic band gap between the low and high frequency regions of optical branches is 0.88 THz. There is no frequency gap between the acoustic and optical branches, which means the presence of strong coupling between acoustic and optical phonons. This optical-acoustic coupling leads to enhanced phonon scattering [1,6,9] and contributes to the low $\kappa_l$. In addition, Fig. 2(f) displays the phonon DOS spectra of BaSnO$_3$, in which the high-frequency region is predominately contributed by O atoms, while Ba and Sn atoms mainly contribute to the low-frequency region. These findings result from the different atom masses between O and Ba (Sn). Meanwhile, ZA and TA modes along the $\Gamma$–$R$ and $\Gamma$–$M$ paths become narrow and almost degenerate, which means the appearance of the strong acoustic-acoustic coupling. Similar coupling phenomenon in acoustic-optical mode has also been observed between LA and the lowest optical branch. These coupling phenomena have also been reported in other perovskite materials and contribute to the low $\kappa_l$, such as CaZrS$_3$ [11].

**3.3 Electrical transport properties**

On the basis of the stable structure, we investigate electrical transport properties to clarify the excellent TE properties in BaSnO3 crystal. The Seebeck coefficients are shown in Fig. 3(a) and 3(b). Obviously, the absolute values of Seebeck coefficients decrease as carrier concentration increases. This indicates that we can obtain the proper $S$ through the carrier concentration strategy. Moreover, it is obvious that the magnitude $S$ increases with temperature. How carrier concentration ($n$) and temperature ($T$) affect the Seebeck coefficient is explained by: [6,44]

$$S = \frac{8\pi^2 k_B^2}{3eh^2} m^* T \left(\frac{\pi}{3n}\right)^{2/3} \tag{1}$$

where the $k_B$, $e$, and $h$ are the Boltzmann constant, electron charge, and Planck constant, respectively. It is clear that the $S$ in Eq (1) is inversely related to $n$ and proportionally to $T$. On the other hand, for identical temperature and carrier concentration, the $n$-type Seebeck coefficient is smaller than that of the $p$-type. This is because the Seebeck coefficient is proportional to the effective DOS (*i.e.*, the slope of DOS) through the following equations:

$$S_p = \frac{k_B}{e} \left[\ln\left(\frac{N_V}{p}\right) + 2.5 - r\right] \tag{2}$$

$$S_n = -\frac{k_B}{e} \left[\ln\left(\frac{N_C}{n}\right) + 2.5 - r\right] \tag{3}$$

where $S_p$, $S_n$, $r$, $N_V$, and $N_C$ are Seebeck coefficients for $p$- and $n$-type doping, the scattering mechanism parameter, the effective DOS for the valence band and conduction band, respectively. As mentioned earlier, from Fig. 1(c) and 1(d), the slope of DOS distribution near the VBM are larger than that of CBM, which thus results in a larger Seebeck coefficient for $p$-type doping. For example, at $T = 300$ K and $n = 3.41 \times 10^{18}$ cm$^{-3}$, the magnitude of $S$ for $p$-type and $n$-type doping are 598 and 194 μV/K, respectively. This means that the *p*-type doping can significantly improve the Seebeck coefficient of BaSnO$_3$.

We now return to the fact that the electrical conductivity ($\sigma$) depends on the carrier relaxation time ($\tau$). Herein, to obtain the electrical conductivity of BaSnO$_3$, we conduct relaxation time calculations on the basis of the deformation potential (DP) theory [38] which was extensively adopted in the TE field [4-6,45]. The relaxation time is given by: [4,6]

$$\tau = \frac{\hbar^4 C_{ii}}{8\pi^3 k_B T (2m^*)^{3/2} E_l^2} \quad (4)$$

where the effective mass *m**, elastic constant $C_{ii}$ and DP constant $E_l$ are defined as $m^*=\hbar^2/(\partial^2 E/\partial k^2)$, $C_{ii}=[\partial^2 E/\partial(\Delta l/l_0)^2]/V_0$, and $E_l=\partial E_{edge}/(\partial(\Delta l/l_0))$, respectively. At the same time, the carrier mobility ($\mu$) can be effectively obtained by:

$$\mu = \frac{\tau e}{m^*} \quad (5)$$

The results of relevant parameters are listed in Table 1. The relaxation time $\tau$ and carrier mobility $\mu$ as a function of temperature for BaSnO$_3$ are shown in Fig. 4(a) and 4(b), respectively. First, the relaxation time and carrier mobility are dependent on carrier types. The differences of *m** between electrons and holes result in different $\tau$ and $\mu$ between *n*- and *p*-type BaSnO$_3$. For example, the relaxation time and carrier mobility of the electron are always higher than that of the hole at identical temperature. Second, the $\tau$ and $\mu$ are dependent on the temperature. Specifically, the relaxation time and carrier mobility decrease with increasing temperature owing to the strong scattering of electron and hole at high temperatures. For example, the relaxation times of *n*- and *p*-type BaSnO$_3$ vary from 24.34 to 8.11 *f*s and 14.90 to 4.97 *f*s at temperatures ranging from 300 to 900 K, respectively. Meanwhile, the carrier mobility of *n*- and *p*-type

BaSnO$_3$ vary from 307.94 to 102.65 cm$^2$ V$^{-1}$ s$^{-1}$ and 19.54 to 6.51 cm$^2$ V$^{-1}$ s$^{-1}$ at the same temperature range, respectively. In fact, the close correlation between relaxation time and temperature is also observed in other TE materials. For instance, the relaxation times of GeSe [46] and GeAs$_2$ [47] are reported to be about 32–4 *f*s and 13–2 *f*s with temperature variation from 300 to 800 K, respectively. Besides, the electron mobility of 307.94 cm$^2$ V$^{-1}$ s$^{-1}$ we obtained (as given in Table 1) is close to the experimental value of 320 cm$^2$ V$^{-1}$ s$^{-1}$ for single crystal at room temperature [50]. In a word, carrier types and temperature nontrivially affect the relaxation time of BaSnO$_3$.

To investigate the power factor, we calculate the electrical conductivity (*σ*) as a function of carrier concentration (Fig. 3(c) and 3(d)) on the basis of the relaxation time. For the identical temperature, the electrical conductivity of both *p*- and *n*-type BaSnO$_3$ increase with carrier concentration. The reason is that electrical conductivity depends on the carrier concentration *n* and carrier mobility *μ* according to the following equation: 3

$$\sigma = n\mu e \tag{6}$$

In addition, the electrical conductivity significantly decreases with increasing temperature, resulting from the strong carrier scattering at high temperatures. This suppresses the transmission of the carrier, which decreases carrier mobility (shown in Fig. 4(b)). The decrease of mobility further reduces the electrical conductivity. Moreover, we find that for the identical carrier concentration and temperature, the *σ* of *p*-type BaSnO$_3$ is significantly lower than that of *n*-type. For instance, at *T* = 300 K and *n* = 3.41×10$^{18}$ cm$^{-3}$, the electrical conductivity of *n*-type BaSnO$_3$ is 12198 S/m, which is significantly higher than that of 822 S/m for *p*-type. These high electrical conductivity of BaSnO$_3$ is comparable to other traditional inorganic perovskites. For example, the highest electrical conductivity of CsSnBr$_3$ [12] and CsSnI$_3$ [12] are reported to be about 1.7 S/cm and 295 S/cm at 300 K, respectively.

The power factor *PF* ($S^2\sigma$) is one of the key parameters for estimating the performance of thermoelectric materials. On the basis of the *S* and $\sigma$, power factors of BaSnO$_3$ that depend on carrier concentration are given in Fig. 3(e) and 3(f), showing that *PF*s first increase and then decrease with the increase of carrier concentration. Therefore, we can obtain the optimal *PF* by balancing the carrier concentration. The reason is that the Seebeck coefficient and electrical conductivity possess the opposite trend with carrier concentration increases. Besides, the *PF*s of *p*-type doped are always greater than those of *n*-type for the identical temperature and carrier concentration based on the large Seebeck coefficient. More importantly, the power factors remain at the highest value of 4.31 mW/mK$^2$ at 300 K for *p*-type BaSnO$_3$, which is greater than that of 0.46 mW/mK$^2$ at the same temperature for *n*-type BaSnO$_3$. These results reveal that the cubic BaSnO$_3$ is a potential TE material, and the *p*-type doping effectively improves its TE performance.

**3.4 Thermal transport properties**

To further elucidate the TE potentialities of BaSnO$_3$, we present the electronic thermal conductivity ($\kappa_e$) and lattice thermal conductivity ($\kappa_l$) in Fig. 5(a)- 5(c). The $\kappa_e$ follows the Wiedemann-Franz law: [1,3,5]

$$\kappa_e = L\sigma T \qquad (7)$$

where $L = 2.45×10^{-8}$ WΩ/K² is the Lorenz constant. As $\kappa_e$ is linearly related to $\sigma$, with varying carrier concentration or temperature, the trend of $\kappa_e$ is similar to that of $\sigma$. The $\kappa_l$ of BaSnO₃ dependent on temperatures is given in Fig. 5(c) by the single-mode relaxation-time approximation (RTA) method, and its values corresponding to the 300, 600, and 900 K are 3.4, 1.9, and 1.3 W/mK, respectively. It is clear that $\kappa_l$ decreases with the increases of temperature, which results from the enhanced phonon-phonon scattering at high temperatures. Under the RTA, $\kappa_l$ is given by: [41,48]

$$\kappa_l = \frac{1}{NV_c} \sum_\lambda C_\lambda v_\lambda \otimes v_\lambda \tau_\lambda \tag{8}$$

in which $V_c$ is the volume of the unit cell, $N$ is the number of unit cells, $C_\lambda$, $v_\lambda$, and $\tau_\lambda$ are the heat capacity, group velocity, and lifetime of the phonon mode λ, respectively. Fig. 5(d) shows the heat capacity at different temperatures. The Debye temperature of BaSnO₃ is about 660 K by the Dulong-Petit high-temperature limit (119 J mol⁻¹ K⁻¹). Fig. 5(e) shows the phonon group velocity dependent on the frequency, and the BaSnO₃ possesses the highest phonon group velocities at the frequency of 9 THz. This results from the steeper phonon spectrum curve at 9 THz. To further understand the origin of $\kappa_l$, we obtain the frequency-dependent phonon lifetime of 300 K through Umklapp scattering or anharmonic three-phonon scattering processes [1,9]. As plotted in Fig. 5(f), black dots and colored backgrounds indicate the phonon modes and phonon mode density, respectively. Owing to the strong phonon scattering, phonon lifetime decreases with increasing frequency, and the longest lifetime (~1.45 ps) mainly occurs in the acoustic mode region. This short phonon lifetime of BaSnO₃ is comparable to other perovskites. For example, the longest phonon lifetime of CaZrSe$_3$ [1], CaZrS$_3$ [11], and BaZrS$_3$ [9] are reported to be about 4 ps, 2.25 ps, and 3 ps at 300 K, respectively.

We next investigate how frequency and phonon mean free path (MFP) affect the cumulative LTC. As seen in Fig. 6(a), approximately 90% of the $\kappa_l$ comes from the low-frequency modes. Specifically, at room temperature (300 K), the contributions of acoustic phonon modes (<3.8 THz) to the $\kappa_l$ is about 25%, while the lower (3.8–13 THz) and higher optical phonon modes contribute ~65% and 10% to $\kappa_l$, respectively. Besides, $\kappa_l$ decreases with the increasing temperature. This decrease of $\kappa_l$ results from the increased phonon–phonon interactions and enhanced phonon scattering at high temperatures. For instance, the $\kappa_l$ is reduced about 50% from 300 to 600 K. Cumulative LTCs with respect to phonon MFP shows the dependence of $\kappa_l$ on BaSnO$_3$ sample size, as shown in Fig. 6(b), which is defined as: [41,48]

$$\kappa(l_{\text{MFP}}) = \frac{1}{N}\sum_\lambda \kappa_\lambda \delta(l_{\text{MFP}} - l_\lambda) \qquad (9)$$

in which the phonon MFP is written as: [41,48]

$$l_\lambda = v_\lambda \tau_\lambda \qquad (10)$$

It is clear that the $\kappa_l$ increases with MFP first and then tends to stable at the limit of MFP. This means that optimizing TE performance is possible by shortening the MFP through nanostructures. For example, at 300 K, the $\kappa_l$ of BaSnO$_3$ can be reduced one half by narrowing down the sample size to 2.1 nm (dotted line in Fig. 6(b)). At room temperature, the MFP limit of BaSnO$_3$ is 80 nm, which is comparable to other perovskites. For example, the MFP limits of CaZrSe$_3$ and CaZrS$_3$ [1,11] are reported to be about 138.1 nm and 73 nm at the same temperature, respectively. Furthermore, the MFP limit decreases with the temperature increases. An example includes that the MFP limits of BaSnO$_3$ vary from 80 to 17 nm at temperatures ranging from 300 to 900 K, due to the enhanced phonon scattering at high temperatures.

**3.5 Figure of merit**

Based on the TE properties mentioned above, we obtain the *ZT* values of BaSnO$_3$ (as shown in Fig. 6(c) and 6(d). The optimal *ZT* of BaSnO$_3$ and corresponding |*S*|, σ, *PF*, and carrier concentration are listed in Table 4. Significantly, the optimal *ZT* value of *n*-type BaSnO$_3$ is always smaller than that of *p*-type for the identical temperatures, due to the large *S* and *PF* of the *p*-type doping system. For example, at 300 K, the maximum *ZT* (*ZT*$_{max}$) of 0.29 for *p*-type doping is significantly higher than that of 0.04 for *n*-type. And the highest *ZT* value of *n*-0.37 and *p*-1.52 for BaSnO$_3$ was obtained at 900K with a carrier concentration of $1.77 \times 10^{19}$ and $7.17 \times 10^{20}$ cm$^{-3}$, respectively. These carrier concentration values satisfy the requirements for the most suitable doping concentration ($10^{19}$−$10^{20}$ cm$^{-3}$), and agree with the reported experimental and theoretical results of BaSnO$_3$ [3,26,51-52].

This remarkable *ZT* value of 1.52 for *p*-type BaSnO$_3$ far exceeds that of traditional inorganic perovskites, *e.g.*, CsGeI$_3$, CsPbI$_3$, and CsSnI$_3$ [49]. Therefore, these results suggest that cubic BaSnO$_3$ is a promising TE material for transparent conducting oxides, and *p*-type doping effectively improves its TE performance.

**4. Conclusions**

Collectively, we investigated the electronic and thermal transport properties of transparent conducting oxide BaSnO$_3$. The lattice constants and band gap we obtained agree with the reported experimental and theoretical results. AIMD simulation and phonon calculation show the stable structure of BaSnO$_3$ perovskite. The combination of excellent electrical and thermal properties contributes to the outstanding Seebeck coefficient and power factor, which gives rise to the *ZT* of *n*-0.37 and *p*-1.52. Our results demonstrate that BaSnO$_3$ is a promising material for high-temperature thermoelectric applications. Strategies of doping and nanostructure open prospects for effectively improving the thermoelectric properties of BaSnO$_3$. It is expected to have worthwhile practical significance for related investigations, and efforts to further optimize the performance of this potential perovskite compound and thermoelectric material system are in progress.

**Conflicts of interest**

There are no conflicts to declare.

**Acknowledgements**

This research was supported by the National Natural Science Foundation of China (grant nos. 62262074 and 11903028), the Scientific Research Foundation of the Department of Education of Yunnan Province, China (2023J1208), and the Special Basic Cooperative Research Programs of Yunnan Provincial Undergraduate Universities' Association (grant NO. 202101BA070001-144). All calculations and simulations were performed with the high-density computing system at the key laboratory of high-density computing, Zhaotong University.

**References**

(1) Osei-Agyemang, E.; Adu, C. E.; Balasubramanian, G. Ultralow lattice thermal conductivity of chalcogenide perovskite CaZrSe$_3$ contributes to high thermoelectric figure of merit. *npj computational materials* **2019**, *5* (116). DOI: 10.1038/s41524-019-0253-5.

(2) Li, M.; Wang, N.; Jiang, M.; Xiao, H.; Zhang, H.; Liu, Z.; Zu, X.; Qiao, L. Improved thermoelectric performance of bilayer Bi$_2$O$_2$Se by the band convergence approach. *Journal of Materials Chemistry C* **2019**, *7* (35), 11029-11039. DOI: 10.1039/C9TC02188D.

(3) Tan, G.; Zhao, L.-D.; Kanatzidis, M. G. Rationally designing high-performance bulk thermoelectric materials. *Chemical Reviews* **2016**, *116* (19), 12123–12149. DOI: 10.1021/acs.chemrev.6b00255.

(4) Wang, N.; Li, M.; Xiao, H.; Gao, Z.; Liu, Z.; Zu, X.; Li, S.; Qiao, L. Band degeneracy enhanced thermoelectric performance in layered oxyselenides by first-principles calculations. *npj computational materials* **2021**, *7* (18). DOI: 10.1038/s41524-020-00476-3.

(5) Wang, N.; Li, M.; Xiao, H.; Gong, H.; Liu, Z.; Zu, X.; Qiao, L. Optimizing the thermoelectric transport properties of Bi$_2$O$_2$Se monolayer via biaxial strain. *Physical Chemistry Chemical Physics* **2019**, *21* (27), 15097-15105. DOI: 10.1039/C9CP02204J.

(6) Wang, N.; Li, M.; Xiao, H.; Zu, X.; Qiao, L. Layered LaCuOSe: A promising anisotropic thermoelectric material. *Physical Review Applied* **2020**, *13* (2), 024038. DOI: 10.1103/PhysRevApplied.13.024038.

(7) Li, J.-F.; Liu, W.-S.; Zhao, L.-D.; Zhou, M. High-performance nanostructured thermoelectric materials. *NPG Asia Materials* **2010**, *2*, 152–158.

(8) Osei-Agyemang, E.; Adu, C. E.; Balasubramanian, G. Doping and anisotropy–dependent electronic transport in chalcogenide perovskite $CaZrSe_3$ for high thermoelectric efficiency. *Advanced Theory Simulations* **2019**, *2* (9), 1900060. DOI: 10.1002/adts.201900060.

(9) Osei-Agyemang, E.; Balasubramanian, G. Understanding the extremely poor lattice thermal transport in chalcogenide perovskite $BaZrS_3$. *ACS Applied Energy Materials* **2020**, *3* (1), 1139–1144. DOI: 10.1021/acsaem.9b02185.

(10) Osei-Agyemang, E.; Koratkarb, N.; Balasubramanian, G. Examining the electron transport in chalcogenide perovskite $BaZrS_3$. *Journal of Materials Chemistry C* **2021**, *9* (11), 3892-3900. DOI: 10.1039/D1TC00374G.

(11) Song, X.; Shai, X.; Deng, S.; Wang, J.; Li, J.; Ma, X.; Li, X.; Wei, T.; Ren, W.; Gao, L.; et al. Anisotropic chalcogenide perovskite $CaZrS_3$: A promising thermoelectric material. *The Journal of Physical Chemistry C* **2022**, *126* (28), 11751–11760. DOI: 10.1021/acs.jpcc.2c02286.

(12) Xie, H.; Hao, S.; Bao, J.; Slade, T. J.; Snyder, G. J.; Wolverton, C.; Kanatzidis, M. G. All-inorganic halide perovskites as potential thermoelectric materials: dynamic cation off-centering induces ultralow thermal conductivity. *Journal of the American Chemical Society* **2020**, *142* (20), 9553–9563.

(13) Grinberg, I.; West, D. V.; Torres, M.; Gou, G.; Stein, D. M.; Wu, L.; Chen, G.; Gallo, E. M.; Akbashev, A. R.; Davies, P. K.; et al. Perovskite oxides for visible-light-absorbing ferroelectric and photovoltaic materials. *Nature* **2013**, *503* (7477), 509–512.

(14) Ramovatar; Coondoo, I.; Kumar, P.; Khan, A. A.; Satapathy, S.; Panwar, N. Observation of large electrocaloric properties in lead-free $Ba_{0.98}Ca_{0.02}Ti_{0.98}Sn_{0.02}O_3$ ceramics. *AIP Advances* **2019**, *9* (5), 055010.

(15) Yin, W.-J.; Weng, B.; Ge, J.; Sun, Q.; Li, Z.; Yan, Y. Oxide perovskites, double perovskites and derivatives for electrocatalysis, photocatalysis, and photovoltaics. *Energy & Environmental Science* **2019**, *12* (2), 442-462.

(16) Arjun, N.; Pan, G.-T.; Yang, T. C. K. The exploration of Lanthanum based perovskites and their complementary electrolytes for the supercapacitor applications. *Results in Physics* **2017**, *7*, 920-926.

(17) Rajasekaran, P.; Arivanandhan, M.; Kumaki, Y.; Jayavel, R.; Hayakawa, Y.; Shimomura, M. Facile synthesis of morphology-controlled La:$BaSnO_3$ for the enhancement of thermoelectric power factor. *CrystEngComm* **2020**, *22* (32), 5363-5374.

(18) Chahib, S.; Fasquelle, D.; Leroy, G. Density functional theory study of structural, electronic and optical properties of cobalt-doped $BaSnO_3$. *Materials Science in Semiconductor Processing* **2022**, *137*, 106220.

(19) Rajasekaran, P.; Kumaki, Y.; Arivanandhan, M.; Khaleeullah, M. M. S. I.; Jayavel, R.; Nakatsugawa, H.; Hayakawa, Y.; Shimomura, M. Effect of Sb substitution on structural, morphological and electrical properties of $BaSnO_3$ for thermoelectric application. *Physica B: Condensed Matter* **2020**, *597*, 412387.

(20) Wei, R.; Tang, X.; Hu, L.; Luo, X.; Yang, J.; Song, W.; Dai, J.; Zhu, X.; Sun, Y. Growth, microstructures, and optoelectronic properties of epitaxial $BaSn_{1-x}Sb_xO_{3-\delta}$ thin films by chemical solution deposition. *ACS Applied Energy Materials* **2018**, *1* (4), 1585–1593.

(21) Scanlon, D. O. Defect engineering of BaSnO$_3$ for high-performance transparent conducting oxide applications. *Physical Review B* **2013**, *87* (16), 161201.

(22) Shin, J.; Kim, Y. M.; Kim, Y.; Park, C.; Char, K. High mobility BaSnO$_3$ films and field effect transistors on non-perovskite MgO substrate. *Applied Physics Letters* **2016**, *109* (26), 262102.

(23) Raghavan, S.; Schumann, T.; Kim, H.; Zhang, J. Y.; Cain, T. A.; Stemmer, S. High-mobility BaSnO$_3$ grown by oxide molecular beam epitaxy. *APL Materials* **2016**, *4* (1), 016106.

(24) Kim, U.; Park, C.; Ha, T.; Kim, Y. M.; Kim, N.; Ju, C.; Park, J.; Yu, J.; Kim, J. H.; Char, K. All-perovskite transparent high mobility field effect using epitaxial BaSnO$_3$ and LaInO$_3$. *APL Materials* **2015**, *3* (3), 036101.

(25) Kim, H. J.; Kim, T. H.; Lee, W.-J.; Chai, Y.; Kim, J. W.; Jwa, Y. J.; Chung, S.; Kim, S. J.; Sohn, E.; Lee, S. M.; et al. Determination of temperature-dependent thermal conductivity of a BaSnO$_{3-\delta}$ single crystal by using the 3ω method. *Thermochimica Acta* **2014**, *585*, 16-20.

(26) Li, J.; Ma, Z.; Sa, R.; Wu, K. Improved thermoelectric power factor and conversion efficiency of perovskite barium stannate. *RSC Advances* **2017**, *7*, 32703-32709.

(27) Spooner, K. B.; Ganose, A. M.; Scanlon, D. O. Assessing the limitations of transparent conducting oxides as thermoelectrics. *Journal of Materials Chemistry A* **2020**, *8*, 11948-11957. DOI: 10.1039/D0TA02247K.

(28) Rajasekaran, P.; Arivanandhan, M.; Sato, N.; Kumaki, Y.; Mori, T.; Hayakawa, Y.; Hayakawa, K.; Kubota, Y.; Jayavel, R.; Shimomura, M. The effect of Sr and Sb co-doping on structural, morphological and thermoelectric properties of $BaSnO_3$ perovskite material. *Journal of Alloys and Compounds* **2022**, *894*, 162335. DOI: 10.1016/j.jallcom.2021.162335.

(29) Balamurugan, K.; Kumar, N. H.; Chelvane, J. A.; Santhosh, P. N. Room temperature ferromagnetism in Fe-doped $BaSnO_3$. *Journal of Alloys and Compounds* **2009**, *472* (1–2), 9-12. DOI: 10.1016/j.jallcom.2008.04.096.

(30) Dahbi, S.; Tahiri, N.; Bounagui, O. E.; Ez-Zahraouy, H. The new eco-friendly lead-free zirconate perovskites doped with chalcogens for solar cells: Ab initio calculations. *Optical Materials* **2020**, *109*, 110442. DOI: 10.1016/j.optmat.2020.110442.

(31) Hinatsu, Y. Electron paramagnetic resonance spectra of $Pr^{4+}$ in $BaCeO_3$, $BaZrO_3$, $BaSnO_3$, and their solid solutions. *Journal of Solid State Chemistry* **1996**, *122* (2), 384-389. DOI: 10.1006/jssc.1996.0131.

(32) Kim, H. J.; Kim, J.; Kim, T. H.; Lee, W.-J.; Jeon, B.-G.; Park, J.-Y.; Choi, W. S.; Jeong, D. W.; Lee, S. H.; Yu, J.; et al. Indications of strong neutral impurity scattering in $Ba(Sn,Sb)O_3$ single crystals. *Physical Review B* **2013**, *88* (12), 125204.

(33) Kresse, G.; Furthmüller, J. Efficient iterative schemes for ab initio total-energy calculations using a plane-wave basis set. *Physical review B* **1996**, *54* (16), 11169-11186. DOI: 10.1103/PhysRevB.54.11169.

(34) Perdew, J. P.; Burke, K.; Ernzerhof, M. Generalized gradient approximation made simple. *Physical Review Letters* **1997**, *77* (18), 3865. DOI: 10.1103/PhysRevLett.77.3865.

(35) Perdew, J. P.; Ruzsinszky, A.; Csonka, G. I.; Vydrov, O. A.; Scuseria, G. E.; Constantin, L. A.; Zhou, X.; Burke, K. Restoring the density-gradient expansion for exchange in solids and surfaces. *Physical Review Letters* **2009**, *100* (13), 136406. DOI: 10.1103/PhysRevLett.100.136406.

(36) Heyd, J.; Scuseria, G. E.; Ernzerhof, M. Hybrid functionals based on a screened Coulomb potential. *The Journal of Chemical Physics* **2003**, *118* (18), 8207. DOI: 10.1063/1.1564060.

(37) Madsen, G. K. H.; Carrete, J.; Verstraete, M. J. BoltzTraP2, a program for interpolating band structures and calculating semi-classical transport coefficients. *Computer Physics Communications* **2018**, *231*, 140-145. DOI: 10.1016/j.cpc.2018.05.010.

(38) Bardeen, J.; Shockley, W. Deformation potentials and mobilities in non-polar crystals. *Physical Review* **1950**, *80* (1), 72-80. DOI: 10.1103/PhysRev.80.72.

(39) Nosé, S. A unified formulation of the constant temperature molecular dynamics methods. *The Journal of Chemical Physics* **1984**, *81* (1), 511. DOI: 10.1063/1.447334.

(40) Togo, A.; Tanaka, I. First principles phonon calculations in materials science. *Scripta Materialia* **2015**, *108*, 1-5. DOI: 10.1016/j.scriptamat.2015.07.021.

(41) Togo, A.; Chaput, L.; Tanaka, I. Distributions of phonon lifetimes in Brillouin zones. *Physical Review B* **2015**, *91* (9), 094306. DOI: 10.1103/PhysRevB.91.094306.

(42) Kim, B. G.; Jo, J. Y.; Cheong, S. W. Hybrid functional calculation of electronic and phonon structure of BaSnO$_3$. *Journal of Solid State Chemistry* **2013**, *197*, 134-138. DOI: 10.1016/j.jssc.2012.08.047.

(43) Krishnaswamy, K.; Himmetoglu, B.; Kang, Y.; Janotti, A.; Walle, C. G. V. d. First-principles analysis of electron transport in BaSnO$_3$. *Physical Review B* **2017**, *95* (20), 205202. DOI: 10.1103/PhysRevB.95.205202.

(44) Guo, D.; Hu, C.; Xi, Y.; Zhang, K. Strain effects to optimize thermoelectric properties of doped Bi$_2$O$_2$Se via tran–blaha modified becke–johnson density functional theory. *The Journal of Physical Chemistry C* **2013**, *117* (41), 21597–21602. DOI: 10.1021/jp4080465.

(45) Fan, Q.; Yang, J.; Qi, H.; Yu, L.; Qin, G.; Sun, Z.; Shen, C.; Wang, N. Anisotropic thermal and electrical transport properties induced high thermoelectric performance in an Ir$_2$Cl$_2$O$_2$ monolayer. *Physical Chemistry Chemical Physics* **2022**, *24* (18), 11268-11277. DOI: 10.1039/D1CP04971B.

(46) Hao, S.; Shi, F.; Dravid, V. P.; Kanatzidis, M. G.; Wolverton, C. Computational prediction of high thermoelectric performance in hole doped layered GeSe. *Chemistry of Materials* **2016**, *28* (9), 3218–3226.

(47) Wang, F. Q.; Guo, Y.; Wang, Q.; Kawazoe, Y.; Jena, P. Exceptional thermoelectric properties of layered GeAs$_2$. *Chemistry of Materials* **2017**, *29* (21), 9300–9307.

(48) Togo, A. First-principles phonon calculations with Phonopy and Phono3py. *Journal of the Physical Society of Japan* **2023**, *92* (1), 012001. DOI: 10.7566/JPSJ.92.012001.

(49) Jong, U.-G.; Kim, Y.-S.; Ri, C.-H.; Kye, Y.-H.; Yu, C.-J. High thermoelectric performance in the cubic inorganic cesium iodide perovskites CsBI$_3$ (B = Pb, Sn, and Ge) from first-principles. *The Journal of Physical Chemistry C* **2021**, *125* (11), 6013–6019.

(50) Kim, H. J.; Kim, U.; Kim, H. M.; Kim, T. H.; Mun, H. S.; Jeon, B.-G.; Hong, K. T.; Lee, W.-J.; Ju, C.; Kimy, K. H.; et al. High mobility in a stable transparent perovskite oxide. *Applied Physics Express* **2012**, *5* (6), 061102. DOI: 10.1143/APEX.5.061102.

(51) Wu, T.; Gao, P. Development of perovskite-type materials for thermoelectric application. *Materials* **2018**, *11* (6), 999. DOI: 10.3390/ma11060999.

(52) Haque, M. A.; Kee, S.; Villalva, D. R.; Ong, W.-L.; Baran, D. Halide perovskites: thermal transport and prospects for thermoelectricity. *Advanced Science* **2020**, *7* (10), 1903389. DOI: 10.1002/advs.201903389.

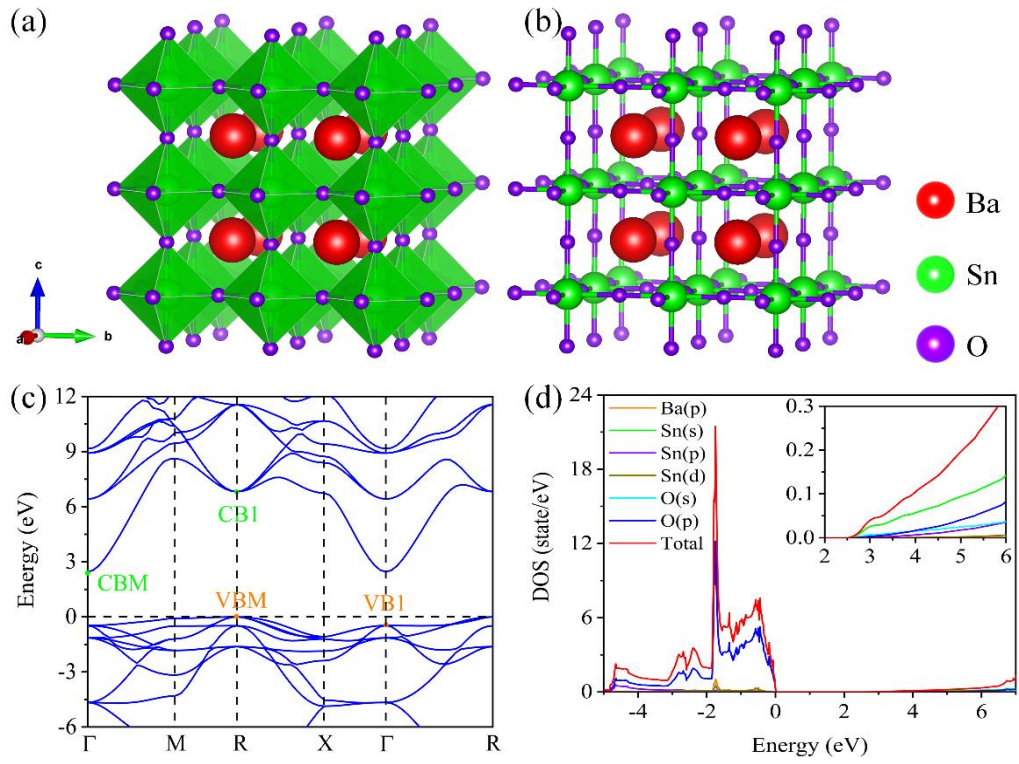

**Fig. 1** The crystal structure (a)-(b), band (c) and density of states (DOS) distribution (d) of BaSnO$_3$.

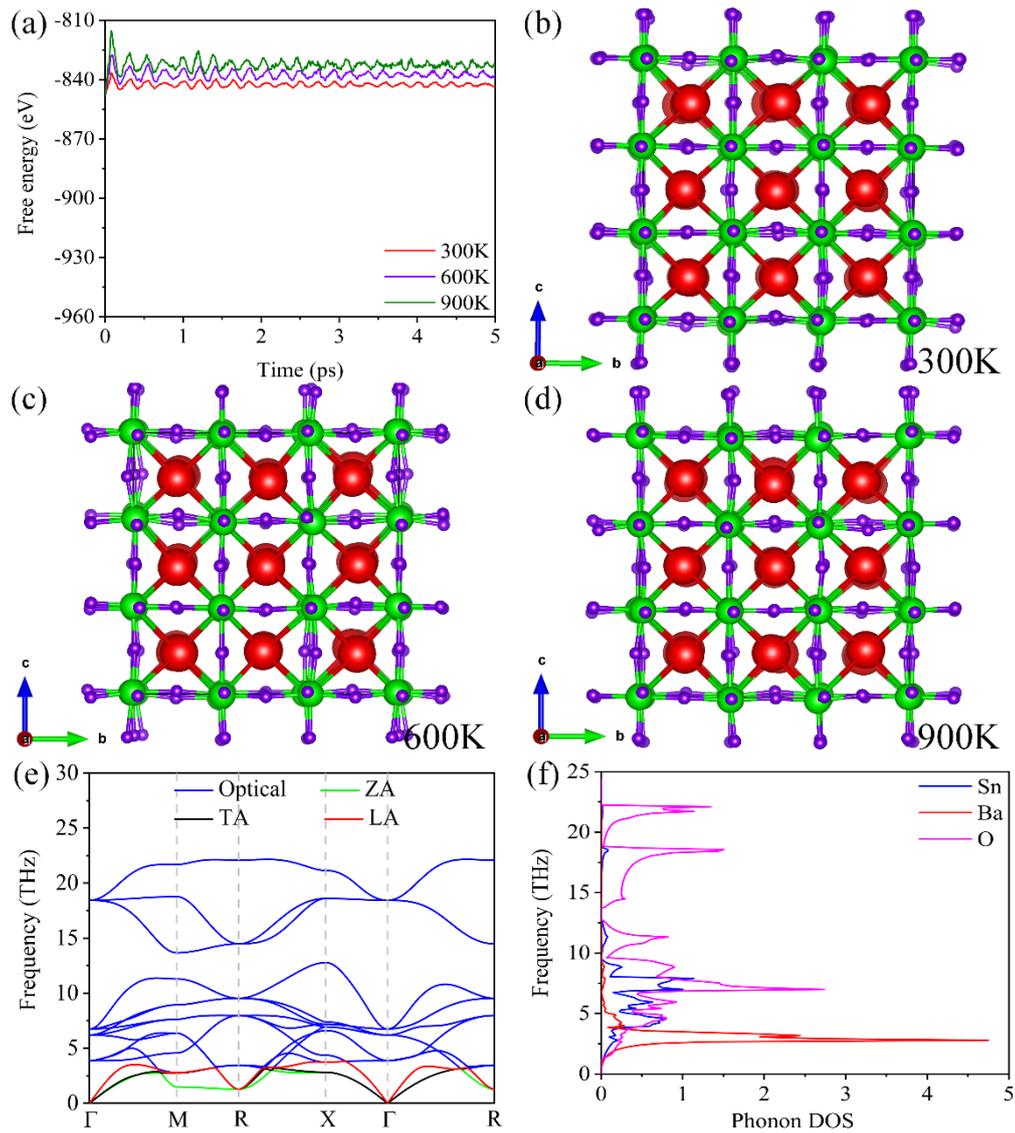

**Fig. 2** Free energy fluctuations with respect to time in AIMD simulations (a) and equilibrium structures of BaSnO$_3$ obtained by AIMD simulations at 300–900 K (b)-(d). The phonon structures (e) and phonon DOS (f) of BaSnO$_3$.

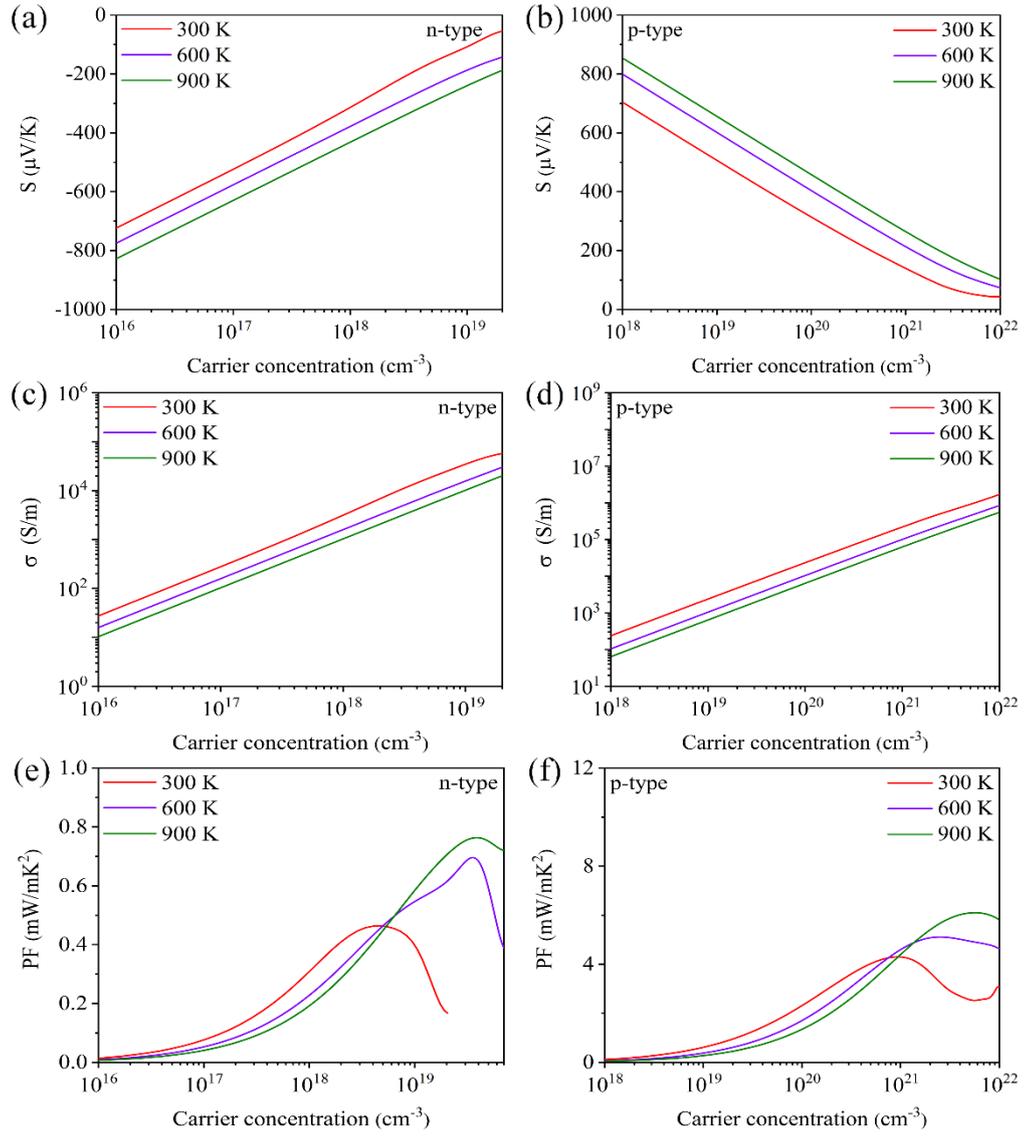

**Fig. 3** The Seebeck coefficient *S* (a)-(b), electrical conductivity *σ* (c)-(d), and power factor *PF* (e)-(f) for BaSnO$_3$ *vs*. carrier concentration.

**Table 1.** The DP constant $E_1$, elastic constant $C$, effective mass $m^*$, relaxation time $\tau$ (300 K), and carrier mobility $\mu$ (300 K) for BaSnO$_3$. The $m_e$ represents the rest mass of the electron.

| Carrier type | $m^*$ ($m_e$) | $C$ (GPa) | $E_l$ (eV) | $\tau$ (fs) | $\mu$ (cm$^2$ V$^{-1}$ s$^{-1}$) |
|---|---|---|---|---|---|
| Electron | 0.14 | 697.51 | 33.39 | 24.34 | 307.94 |
| Hole | 1.34 | 697.51 | 7.80 | 14.90 | 19.54 |

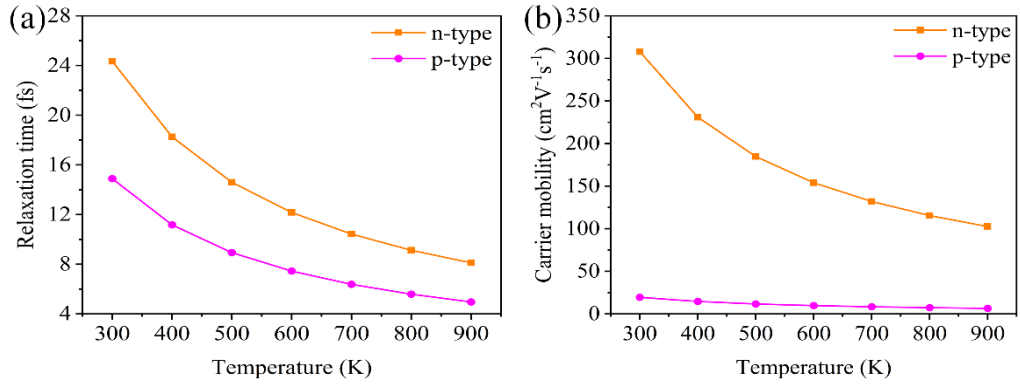

**Fig. 4** Relaxation time $\tau$ and carrier mobility $\mu$ as a function of temperature for BaSnO$_3$.

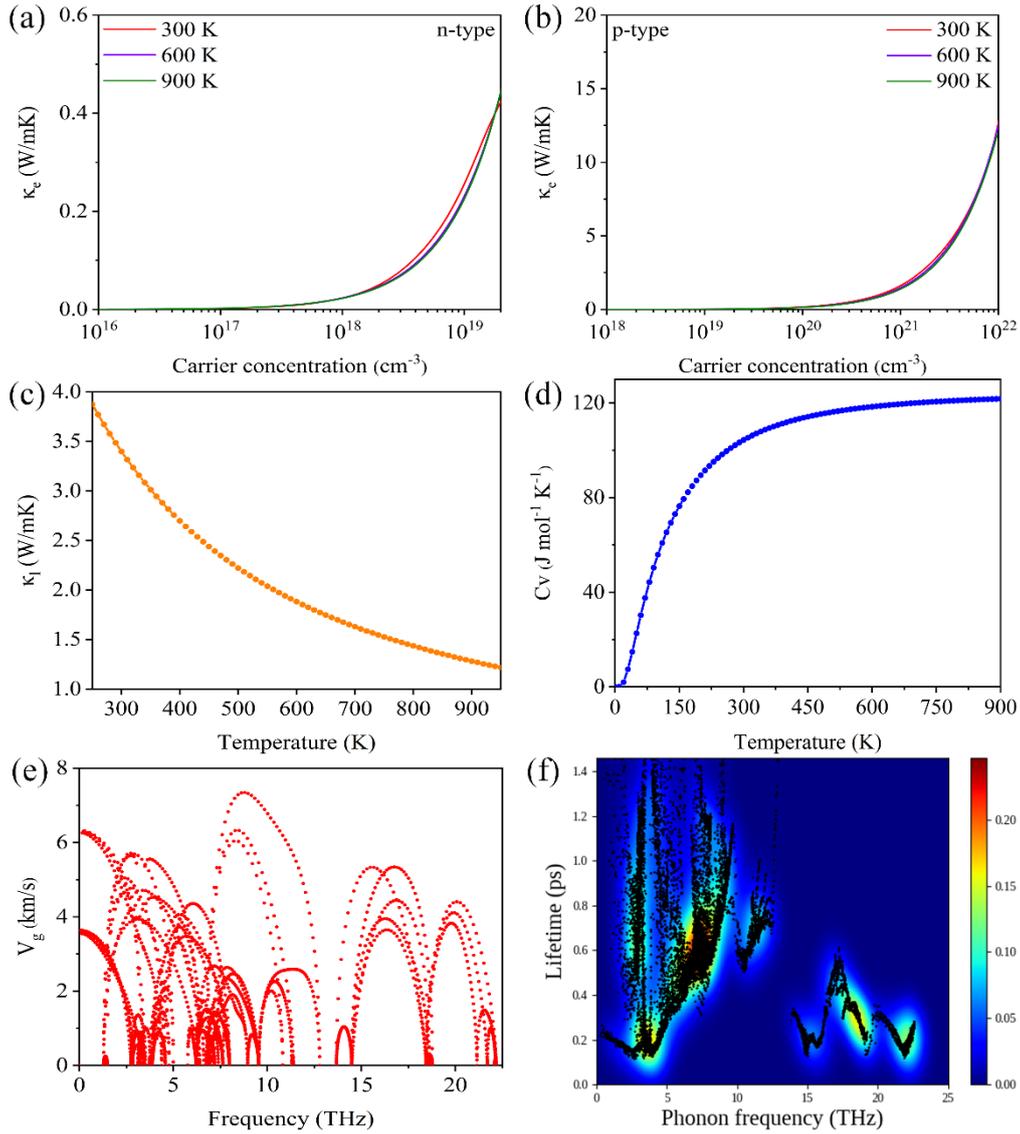

**Fig. 5** The electronic thermal conductivity $\kappa_e$ (a)-(b) of BaSnO$_3$ as a function of carrier concentration. The lattice thermal conductivity $\kappa_l$ (c) and heat capacity $C_v$ (d) *vs.* temperature. Phonon group velocity $V_g$ (e) and phonon lifetime (f) *vs.* phonon frequency at 300 K for BaSnO$_3$.

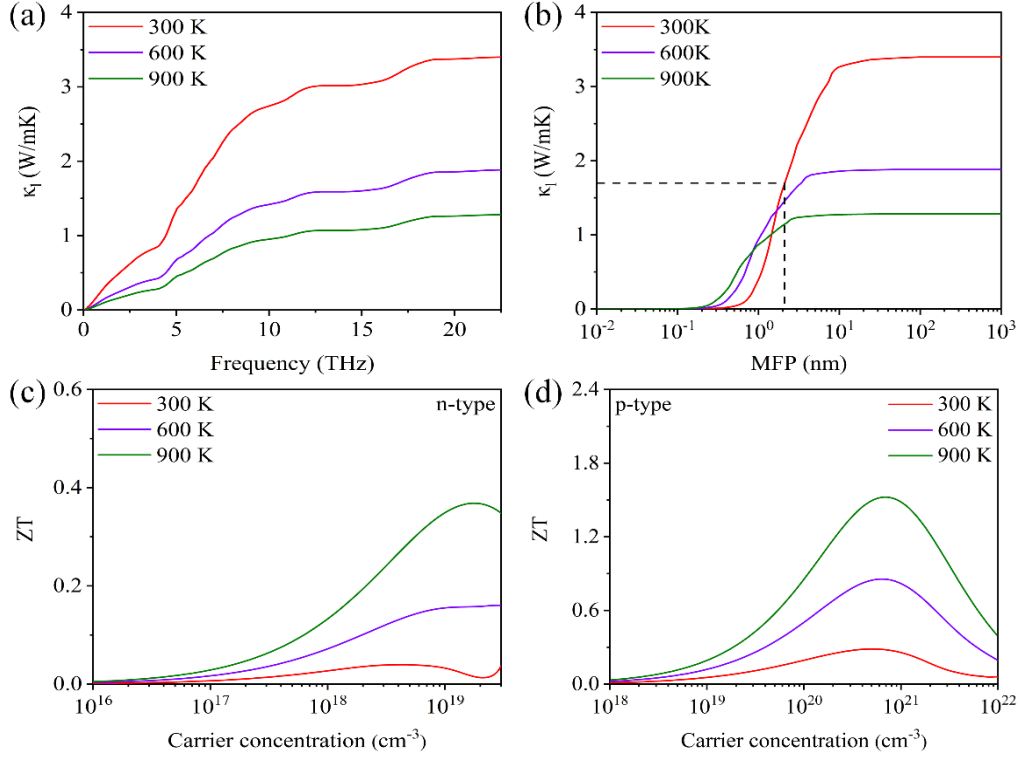

**Fig. 6** The cumulative lattice thermal conductivity as a function of phonon frequency (a) and phonon mean free path (b) at 300, 600, and 900 K for BaSnO$_3$. The figure of merit (*ZT*) as a function of carrier concentration of *n*-type (c) and *p*-type (d) BaSnO$_3$.

**Table 4.** Optimal *ZT* values and corresponding parameters conditions for BaSnO$_3$.

| Temperature (K) | carrier type | $n$ ($10^{18}$ cm$^{-3}$) | $|S|$ ($\mu$VK$^{-1}$) | $\sigma$ ($10^3$ Sm$^{-1}$) | PF (mWm$^{-1}$K$^{-2}$) | ZT |
|---|---|---|---|---|---|---|
| 300 | hole | 473.30 | 192.98 | 107.11 | 3.99 | 0.29 |
|  | electron | 4.03 | 178.74 | 14.50 | 0.463 | 0.04 |
| 600 | hole | 636.72 | 249.32 | 65.21 | 4.05 | 0.86 |
|  | electron | 27.70 | 126.33 | 41.88 | 0.6684 | 0.16 |
| 900 | hole | 717.41 | 291.60 | 45.49 | 3.87 | 1.52 |
|  | electron | 17.73 | 195.55 | 17.96 | 0.6869 | 0.37 |